# Modulated Pulses Based High Spatial Resolution Distributed Fiber System for Multi-Parameter Sensing

**JINGDONG ZHANG,[1] TAO ZHU,[1*] HUAN ZHOU,[1] YANG LI,[1] MIN LIU,[1] WEI HUANG[1]**

[1] *Key Laboratory of Optoelectronic Technology & Systems (Education Ministry of China) Chongqing University, Chongqing 400044, China*
*\*zhutao@cqu.edu.cn*

**Abstract:** We demonstrate a hybrid distributed fiber sensing system for multi-parameter detection. The integration of phase-sensitive optical time domain reflectometry (Φ-OTDR) and Brillouin optical time domain reflectometry (B-OTDR) enables measurement of vibration, temperature and strain. Exploiting the fast changing property of vibration and the static property of temperature and strain, the laser pulse width and intensity are modulated and then injected into the single-mode sensing fiber proportionally, so that the three concerned parameters can be extracted simultaneously by only one photo-detector and data acquisition channel. Combining with advanced data processing methods, the modulation of laser pulse brings additional advantages because of trade and balance between the backscattering light power and nonlinear effect noise, which enhances the signal-to-noise ratio, and enables sub-meter level spatial resolution together with long sensing distance. The proposed method realizes up to 4.8 kHz vibration sensing with 3 m spatial resolution at 10 km standard single-mode fiber. And measurements of the distributed temperature and stress profile along the same fiber with 80 cm spatial resolution are also achieved concurrently.

**References and links**

1. J. C. Juarez and H. F. Taylor, "Polarization discrimination in a phase-sensitive optical time-domain reflectometer intrusion-sensor system," Opt. Lett. **30**(24), 3284-3286 (2005).
2. F. Peng, H. Wu, X. Jia, Y. Rao, Z. Wang and Z. Peng, "Ultra-long high-sensitivity Φ-OTDR for high spatial resolution intrusion detection of pipelines," Opt. Express **22**(11), 13804-13810 (2014).
3. M. Jones, "Structural-health monitoring: a sensitive issue," Nat. Photonics **2**(3), 153-154 (2008).
4. L. Thévenaz, "Review and Progress on Distributed Fibre Sensing," in *Optical Fiber Sensors*, OSA Technical Digest (CD) (Optical Society of America, 2006), paper ThC1.
5. X. Bao and L. Chen, "Recent progress in distributed fiber optic sensors," Sensors **12**(7), 8601-8639 (2012).
6. Y. Lu, T. Zhu, L. Chen, and X. Bao, "Distributed vibration sensor based on coherent detection of phase-OTDR," J. Lightwave Technol. **28**(22), 3243-3249 (2010).
7. A. Masoudi, M. Belal, and T. Newson, "A distributed optical fibre dynamic strain sensor based on phase-OTDR," Meas. Sci. Technol. **24**(8), 085204 (2013).
8. T. Zhu, Q. He, X. Xiao, and X. Bao, "Modulated pulses based distributed vibration sensing with high frequency response and spatial resolution," Opt. Express **21**(3), 2953-2963 (2013).
9. J. Pastor-Graells, H. Martins, A. Garcia-Ruiz, S. Martin-Lopez, and M. Gonzalez-Herraez, "Single-shot distributed temperature and strain tracking using direct detection phase-sensitive OTDR with chirped pulses," Opt. Express **24**(12), 13121-13133 (2016).
10. F. Wang, W. Zhan, X. Zhang, and Y. Lu, "Improvement of spatial resolution for BOTDR by iterative subdivision method," J. Lightwave Technol. **31**(23), 3663-3667 (2013).
11. Y. Weng, E. Ip, Z. Pan, and T. Wang, "Single-end simultaneous temperature and strain sensing techniques based on Brillouin optical time domain reflectometry in few-mode fibers," Opt. Express **23**(7), 9024-9039 (2015).
12. Q. Li, J. Gan, Y. Wu, Z. Zhang, J. Li, and Z. Yang, "High Spatial Resolution BOTDR Based on Differential Brillouin Spectrum Technique," IEEE Photonic. Tech. L. **28**(14), 1493-1496 (2016).
13. S. Matsuura, M. Kumoda, Y. Anzai, and Y. Koyamada, "Distributed strain measurement in GI fiber with sub-meter spatial resolution by DP-BOTDR," in *2013 18th OptoElectronics and Communications Conference held jointly with 2013 International Conference on Photonics in Switching*, (Optical Society of America, 2013), paper MS2_2.
14. M. A. Farahani and T. Gogolla, "Spontaneous Raman scattering in optical fibers with modulated probe light for distributed temperature Raman remote sensing," J. Lightwave Technol. **17**(8), 1379-1391 (1999).

15. J. Park, G. Bolognini, D. Lee, P. Kim, P. Cho, F. Di Pasquale, and N. Park, "Raman-based distributed temperature sensor with simplex coding and link optimization," IEEE Photonic. Tech. L. **18**(17), 1879-1881 (2006).
16. M. A. Soto, A. Signorini, T. Nannipieri, S. Faralli, G. Bolognini, and F. Di Pasquale, "Impact of Loss Variations on Double-Ended Distributed Temperature Sensors Based on Raman Anti-Stokes Signal Only," J. Lightwave Technol. **30**(8), 1215-1222 (2012).
17. G. Bolognini and A. Hartog, "Raman-based fibre sensors: Trends and applications," Opt. Fiber Technol. **19**(6), 678-688 (2013).
18. M. Alahbabi, Y. Cho, and T. Newson, "Simultaneous temperature and strain measurement with combined spontaneous Raman and Brillouin scattering," Opt. Lett. **30**(11), 1276-1278 (2005).
19. G. Bolognini and M. A. Soto, "Optical pulse coding in hybrid distributed sensing based on Raman and Brillouin scattering employing Fabry–Perot lasers," Opt. Express **18**(8), 8459-8465 (2010).
20. M. Taki, A. Signorini, C. J. Oton, T. Nannipieri, and F. Di Pasquale, "Hybrid Raman/Brillouin-optical-time-domain-analysis-distributed optical fiber sensors based on cyclic pulse coding," Opt. Lett. **38**(20), 4162-4165 (2013).
21. Y. S. Muanenda, M. Taki, T. Nannipieri, A. Signorini, C. J. Oton, F. Zaidi, I. Toccafondo, and F. Di Pasquale, "Advanced Coding Techniques for Long-Range Raman/BOTDA Distributed Strain and Temperature Measurements," J. Lightwave Technol. **34**(2), 342-350 (2016).
22. R. Bernini, A. Minardo, and L. Zeni, "Dynamic strain measurement in optical fibers by stimulated Brillouin scattering," Opt. Lett. **34**(17), 2613-2615 (2009).
23. Y. Peled, A. Motil, L. Yaron, and M. Tur, "Slope-assisted fast distributed sensing in optical fibers with arbitrary Brillouin profile," Opt. Express **19**(21), 19845-19854 (2011).
24. A. Bergman, L. Yaron, T. Langer, and M. Tur, "Dynamic and distributed slope-assisted fiber strain sensing based on optical time-domain analysis of Brillouin dynamic gratings," J. Lightwave Technol. **33**(12), 2611-2616 (2015).
25. A. Motil, O. Danon, Y. Peled, and M. Tur, "Pump-power-independent double slope-assisted distributed and fast Brillouin fiber-optic sensor," IEEE Photonic. Tech. L. **26**(8), 797-800 (2014).
26. D. Ba, B. Wang, D. Zhou, M. Yin, Y. Dong, H. Li, Z. Lu, and Z. Fan, "Distributed measurement of dynamic strain based on multi-slope assisted fast BOTDA," Opt. Express **24**(9), 9781-9793 (2016).
27. J. Hu, L. Xia, L. Yang, W. Quan, and X. Zhang, "Strain-induced vibration and temperature sensing BOTDA system combined frequency sweeping and slope-assisted techniques," Opt. Express **24**(12), 13610-13620 (2016).
28. Y. Muanenda, C. J. Oton, S. Faralli, T. Nannipieri, A. Signorini, and F. Di Pasquale, "A distributed acoustic and temperature sensor using a commercial off-the-shelf DFB laser," Proc. SPIE **9634**, 96342C(2015)
29. Y. Muanenda, C. J. Oton, S. Faralli, T. Nannipieri, A. Signorini, and F. Di Pasquale, "Hybrid distributed acoustic and temperature sensor using a commercial off-the-shelf DFB laser and direct detection," Opt. Lett. **41**(3), 587-590 (2016).
30. G. P. Agrawal, *Nonlinear fiber optics* (Academic 2007).
31. Y. Hao, Q. Ye, Z. Pan, F. Yang, H. Cai, R. Qu, Q. Zhang, and Z. Yang, "Design of wide-band frequency shift technology by using compact Brillouin fiber laser for Brillouin optical time domain reflectometry sensing system," IEEE Photonics Journal **4**(5), 1686-1692 (2012).
32. Y. Yao, Y. Lu, X. Zhang, F. Wang, and R. Wang, "Reducing trade-off between spatial resolution and frequency accuracy in BOTDR using cohen's class signal processing method," IEEE Photonic. Tech. L. **24**(15), 1337-1339 (2012).
33. S. Huang, T. Zhu, Z. Cao, M. Liu, M. Deng, J. Liu, and X. Li, "Laser Linewidth Measurement Based on Amplitude Difference Comparison of Coherent Envelope," IEEE Photonic. Tech. L. **28**(7), 759-762 (2016).

## 1. Introduction

During the past several decades, many high efficient and flexible distributed fiber sensing systems have been widely used in numerous fields, such as intrusion monitoring [1], oil-gas depot [2], aerospace and wind-energy industries [3]. Among these techniques, Rayleigh, Brillouin or Raman backscattering based optical time domain reflectometer are very practical owing to the advantages of single-ended and simple construction [4,5]. Conventionally, a phase-sensitive optical time-domain reflectometry (Φ-OTDR), exploiting Rayleigh backscattering signal of narrow-linewidth laser pulse, is used to measure weak and fast changing perturbations along critical infrastructures [6-9]. Besides, the Brillouin optical time domain reflectometry (B-OTDR), measuring frequency shift of Brillouin backscattering, has recently been highly concerned for determining distributed temperature and strain over long distances [10-13]. The Raman optical time domain reflectometry (R-OTDR) or Raman distributed temperature sensor (RDTS) [14-17], exploiting avalanche photodetectors (APDs) to detect the different of Raman anti-Stokes (AS) and Stokes (S) signal induced by temperature, is widely used to measure distributed temperature with meter-scale spatial resolutions.

At the meantime, multi-parameter measurement is becoming a pressing requirement for fiber optic monitoring systems because it provides more valuable information which enables a comprehensive identification for fault event. Several novel methods for simultaneous distributed measurement of temperature and strain were reported based on spatially resolving both spontaneous Raman and Brillouin backscattered anti-Stokes signals [18-21]. These methods eliminate the Brillouin frequency shift (BFS) cross-sensitivity between temperature and static strain. Slope-assisted method, based on stimulated Brillouin scattering interaction between two counter propagating optical pulses, permits dynamic strain or low frequency vibration to be measured by Brillouin optical time-domain analysis (BOTDA) schemes [22-27]. For the purpose of sensing temperature and strain-induced vibration simultaneously, a BOTDA scheme combined frequency sweeping and slope-assisted techniques was proposed [27]. Two vibration events and one temperature point are clearly identified. However, slope-assisted method is only suitable for strain-induced low frequency vibration sensing, and it is difficult to measure weak perturbations with $n\varepsilon$-level strain. Recently, a hybrid distributed acoustic and temperature sensor was reported based on integration of RDTS and Φ-OTDR utilizing a commercial off-the-shelf distributed feedback laser[28,29]. It controlled and modulated the optical source to ensure inter-pulse incoherence and intra-pulse coherence, allowing cyclic simplex coding to be effective for both φ-OTDR and RDTS. This scheme enables accurate long-distance measurement of vibrations and temperature with minimal post-processing, which is incapable for static strain measuring.

In this paper, we propose and experimentally demonstrate a hybrid distributed multi-parameter fiber sensing system based on integration of a Φ-OTDR and a B-OTDR. Both the width and intensity of the laser pulse are modulated so that the information of vibration, temperature and strain can be extracted simultaneously by only one photo-detector and data acquisition channel. In particular, a Φ-OTDR system based on direct detection for fast measurement of coherent Raleigh scattering is used with a heterodyne detection of Brillouin Stokes signal and Brillouin fiber laser by the control of an optic switcher (OS). The balances between the backscattering light power and nonlinear effect noise, combining with advanced data processing methods, enable high spatial resolution measurement of vibration, temperature and strain with long-distance.

## 2. Principles

### 2.1 Laser pulse modulation

In order to enhance the spatial resolution and achieve simultaneous multi-parameter measurement in long sensing range, the pulse width and intensity of the laser are appropriated modulated. As shown in Fig. 1, each cycle of the modulated pulses consists of a group of wide pulses $I_1$ with high intensity and a narrow pulse $I_2$ with low intensity. The pulses pattern of single cycle can be written as:

$$y(t) = \sum_{a=1}^{N} I_1(aT - t) + I_2((N+1)T - t) \tag{1}$$

where $I_1(t)$ and $I_2(t)$, the profile of pulses, are Gaussian shape in our system to maintain a balance between pulse power and spatial resolution $I_i(t)$ can be written as:

$$I_i(t) = P_i e^{-t^2 \left/ \left(\frac{\tau_i}{2\sqrt{2\ln 2}}\right)^2 \right.} \tag{2}$$

Herein $P_i$ and $\tau_i$ are peak power and full width at half maximum (FWHM) of the pulse, respectively, with $P_1 > P_2$ and $\tau_1 > \tau_2$.

Vibration is a rapidly changing parameter which can be detected through intensity variation of Rayleigh backscattering in Φ-OTDR, while the static parameters, such as temperature and strain can be detected through Brillouin backscattering frequency shift from B-OTDR. Restricted by the photodetector duck current noise and limited responsivity, the pulses $I_1$ possessing wide duration and high intensity are utilized for vibration detection, due to the fact that intensity based Φ-OTDR needs more optical energy to activate enough Rayleigh backscattering signal in the sensing fiber. Furthermore, according to the Nyquist–Shannon sampling theorem [6,8], sampling frequency should be as high as possible to widen system frequency response range, i.e., the number of pulses for Φ-OTDR sensing should be guaranteed as large as possible. The narrow and low intensity pulse $I_2$ is used for temperature and strain sensing for the reason that frequency based B-OTDR is immune to the duck current noise. However, an high peak pulse will not increase the signal-to-noise ratio (SNR) of B-OTDR, but lead to stimulated Brillouin scattering and other nonlinear optical effects, which will deteriorate SNR instead.

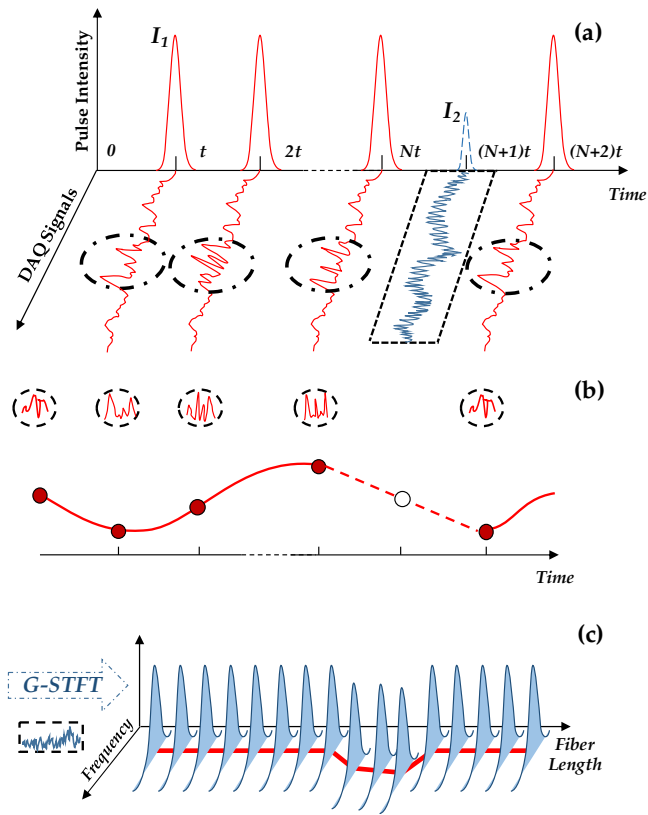

Fig. 1. Laser pulse modulation and data processing. (a) The modulated laser pulse sequences and signals captured by DAQ. (b) φ-OTDR data processing at the vibration point. (c) B-OTDR data processing, data after Gaussian window short time Fourier transform (G-STFT).

To guarantee the coherence of backscattering light in long sensing fiber, a narrow linewidth laser source is a crucial element in Φ-OTDR. However, a narrow linewidth laser with high power often brings significant nonlinear effects [30], such as unstable modulation, stimulated Brillouin, which leads to the randomization of the scattering light the pulse energy transferring from laser frequency to nearby frequencies. These nonlinear effects will deteriorate the system SNR or even make the signal unable to detect. On the other hand, because of fiber attenuation, the far end backscattering light of the long fiber will submerged in the dark current noise of the photo-detector when the peak power is not high enough. Thus this also limits the sensing range.

## 2.2 Principle of operation & data processing

When the sensing fiber experiences vibration, the refractive index and length of the fiber near the local vibrations changes, which ultimately changes the local amplitude and phase of the modulated pulses, and also the intensity of the backscattered signal [29]. Fig. 1(a) and 1(b) show the data processing of distributed vibration measurement. The vibration points can be accurately located by comparing different period of backscattering traces. The (N+1)$^{th}$ trace is utilized to B-OTDR sensing, and the vibration points are vacant but could be estimated by the linear interpolation of the adjacent signals.

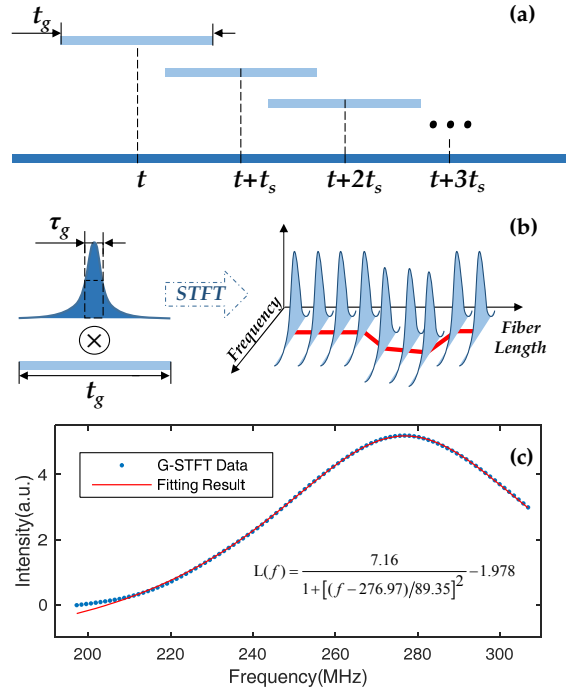

Fig. 2 data processing for B-OTDR. (a) Beat signal of LO and backscattering light is divide into shorter segments of equal length. (b) The processing of Gaussian window short time Fourier transform (G-STFT). (c) Fitting result of weighted Lorentzian nonlinear fitting (WLF)

When experiencing temperature change or strain, linearly red shift happens to the Brillouin scattering frequency shift $v_B$ according to the formula [31]

$$v_B = C_{v_B T} \cdot \Delta T + C_{v_B \varepsilon} \cdot \Delta \varepsilon + v_{B0} \qquad (3)$$

Here $C_{v_B T}$ (~1MHz/°C) and $C_{v_B \varepsilon}$ (~1MHz/20με) are the temperature and strain coefficients for BFS in single mode silica fibers, and $\Delta T$ and $\Delta \varepsilon$ are temperature and strain variation of sensing point, respectively. $v_{B0}$ is Brillouin scattering frequency shift (typical in the range of ~11GHz) of fiber. A stable single-frequency Brillouin fiber laser (BFL), pumped by the narrow-linewidth laser and shifted about hundreds of MHz from $v_B$, is regarded as local oscillator (LO) for coherent detection in this BOTDR sensing system [31]. The beating signal of Brillouin scattering (hundreds of MHz) and LO is digitized and recorded directly by a high sampling rate data acquisition system. Gaussian window short time Fourier transform (G-STFT) method and weighted Lorentzian nonlinear fitting (WLF) method are proceeded to demodulate $v_B$ from the record time domain signal, as shown in Fig. 2.

To get the frequency spectrum at *t*, the beat signal is firstly multiplied with $t_g$ length Gaussian window, and then Fourier transform is taken, which is illustrated in Fig. 2(a). The G-STFT procedure can explained by formula

$$P_{STFT}(t,\omega) = \left| \int_{-t_g/2}^{-t_g/2} e^{-j\omega\tau} s(\tau) g(\tau-t) d\tau \right|^2 \quad (4)$$

where *g(t)* is Gaussian window function

$$g(t) = e^{-t^2 / \left(\frac{\tau_g}{2\sqrt{2\ln 2}}\right)^2} \quad (5)$$

$\tau_g$ is the FWHM of Gaussian window, as shown in Fig. 2(b).

Depending on the Heisenberg-Gabor uncertainty principle [32], the product of the windowed signal duration and the spectrum bandwidth has to be not smaller than a constant, i.e. spatial resolution and frequency bandwidth is a trade-off question in our system. Fortunately, the gain spectrum of spontaneous Brillouin is a single peak Lorentz line-shape, whose bandwidth is tens of MHz normally. By increasing the Gaussian window length $t_g$ while keeping the FWHM $\tau_g$ stable, high spatial resolution and enough frequency bandwidth could be achieved simultaneously with hundreds of sampling periods averaging. Spatial resolution of the OTDR system could be expressed as $\delta_z = c\tau_p/2n_c$, where *c* represents speed of light in vacuum, and $\tau_p$ is laser pulse duration and $n_c$ stands for refractive index of fiber core. Therefore, FWHM of the window function $\tau_g$ is set to be slightly smaller than the pulse width $\tau_p$. Meanwhile, the window width $t_g$ is determined by the DAQ sampling rate and the detected width of spontaneous Brillouin gain spectrum, which $t_g$ is supposed to be $t_g = 12\tau_g$ in the proposed system. In addition, the moving step $t_s$ of Gaussian window is set to $\tau_p/5$ for a better analytical spatial resolution. Then, the frequency spectrum of STFT at position t can be obtained after hundreds of periods average.

After acquiring the STFT frequency spectrum, Lorentzian nonlinear fitting (WLF) method is adopted to get the accurate BFS center frequency. Considering the spectral asymmetry resulted from noises from spontaneous Brillouin gain distortion in practice, the data at bottom right of spectrum peak is not used in curve fitting. What's more, normalized under-fit data, severed as weight value, is expected to multiply with under-fit data during each fitting iteration, which highlights the weight of data near the peak and reduces iterate times. Fitting result is displayed in Fig. 2(c).

## 3. Experiment and discussions

### 3.1 Experimental setup

The experimental setup used to evaluate the performance of the proposed system is shown in Fig. 3(a). A single frequency light source whose linewidth is less than 200Hz (NKT Laser, E15) is injected into the sensing fiber and a Brillouin laser cavity through a coupler. An acoustic-optic modulator (AOM), driven by the arbitrary waveform generator (AWG), is used to generate modulated pulses with different widths and amplitudes for φ-OTDR and B-OTDR. The pulse widths for φ-OTDR and B-OTDR are 30 ns and 8 ns, respectively. Then the light signal is amplified by an Erbium-doped fiber amplifier (EDFA) and a narrowband band-pass filter (BPF) is used to filter Amplified Spontaneous Emission (ASE) noise. Through a circulator, Rayleigh and Brillouin backscattering traces of modulated pulses are amplified by another EDFA and the Brillouin backscattering is selectively beat with Brillouin laser by an optical switch (OS).

To find a balance between signal power and nonlinear effects, we observe the scattering spectra of different pulse power along the 10km sensing fiber. The CW light is modulated to a 9.7 kHz repeating rate light pulse by an AOM. Scattering spectra of sensing fiber at different EDFA pump power is shown in Fig.3(b). There is only Rayleigh and Brillouin scattering peaks when the pump current changes from 70mA to 80mA, and the scattering power increases with pump current. However, when the EDFA pump current goes greater than 80mA, new frequency component generates on the both sides of the light source, while Rayleigh and Brillouin scattering no longer increases. Therefore, the 80mA pump current is chosen in the experiment and the peak power of $I_1$ and $I_2$ are 14W and 3W respectively, as shown in Fig. 3(c).

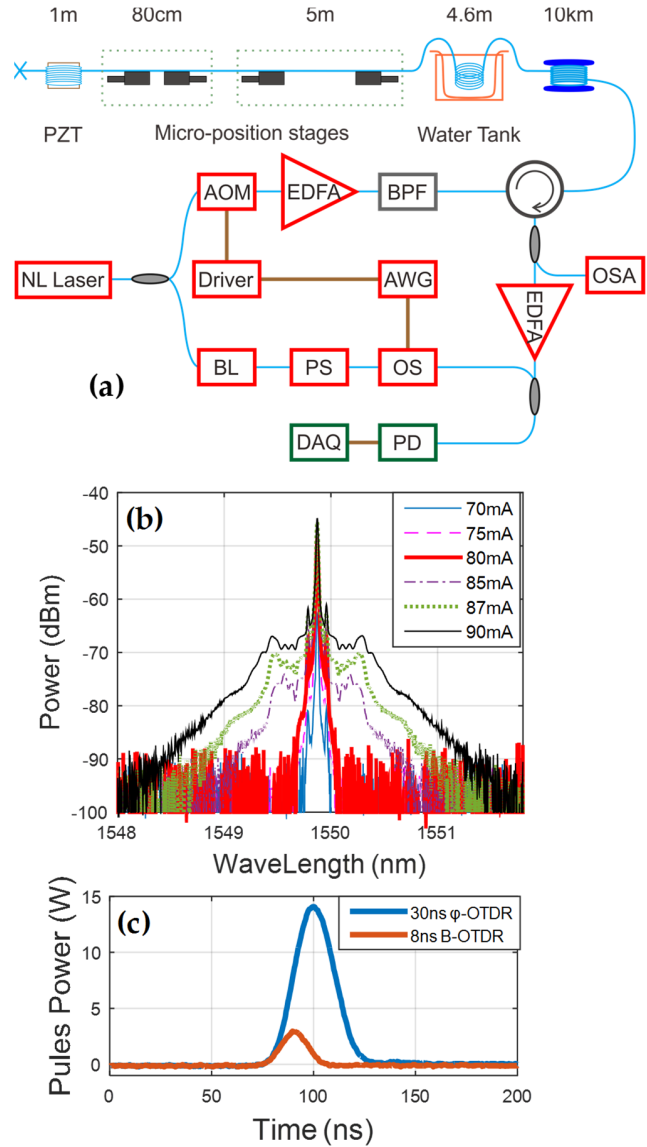

Fig.3 (a) Experimental setup. AOM: Acousto-Optic Modulator, EDFA: Erbium-Doped Fiber Amplifier, BPF: Band-Pass Filter, PZT: Piezoelectric Transducer, AWG: Arbitrary Waveform Generator, BL: Brillouin Laser, PS: Polarization Scrambler, OS: Optical Switch, PD: Photo Detector, DAQ: Data Acquisition card. (b) Scattering spectra of sensing fiber at different EDFA pump power. (c) Amplified optical pulses for B-OTDR and φ-OTDR

At the end of 10170m sensing fiber, about 4.6m fiber are placed in a water tank, two short sections (5m and 80cm) of the fiber are glued on two pairs of micro-position stages to apply the strain, and 1m fiber is wound on a piezoelectric transducer (PZT) tube to simulate external perturbations.

The proposed Brillouin laser cavity is shown in Fig. 4(a), which consists of a EDFA with 25dBm saturation output, a three-port circulator, a polarization controller, 50m scattering fiber and a 10:90 coupler. Fixed in polyurethane foam, the laser cavity is resistant to vibrational perturbation. The scattering fiber here is panda polarization-maintaining fiber whose core and cladding diameter are 8um and 94um, respectively. Heterodyned with the 1550nm narrow linewidth seed laser, the Brillouin frequency shift (BFS) of the Brillouin laser is 10.324GHz as shown in Fig. 4(b). The Side mode suppression ratio is 73 dB and interval of longitudinal mode is about 3.8 MHz which corresponding to 52 m cavity length. The linewidth of Brillouin cavity is ~ 230Hz, measured by the method in [33], as shown in Fig.4(c).

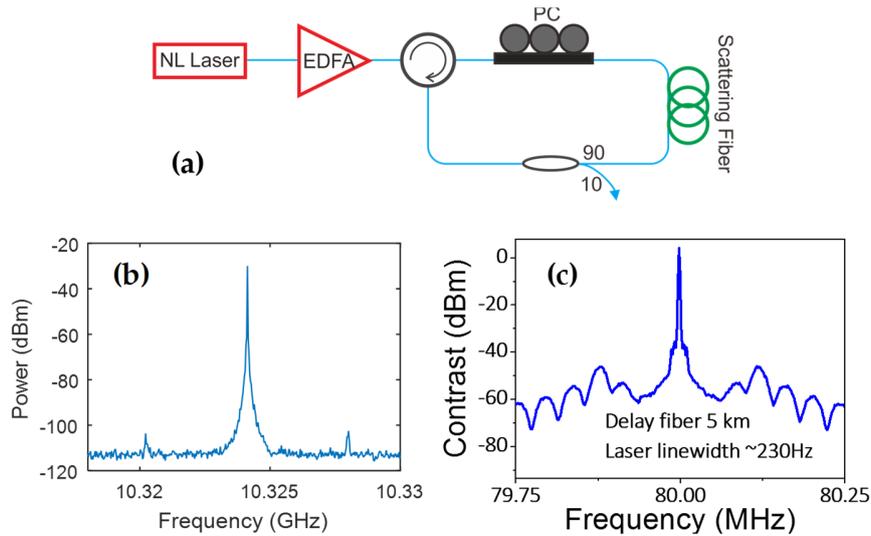

Fig. 4 (a) Brillouin Laser. (b) Beat signal between the NKT laser and the Brillouin Laser. (c)Linewidth measurement spectrum.

In order to simultaneously detect vibration, temperature and strain along the sensing fiber, the 30ns pulses and 8ns pulse with repeating rate of 9.7 KHz are injected into the fiber. As demonstrated before, the ratio of wide pulse number and narrow pulse number is 100:1. That is to say, N in Fig. 1(a) equals 100. To implement φ-OTDR sensing, the OS is switched off after every high power pulses enter into the sensing fiber. While, when OS is switched on, the Stokes Brillouin spontaneous peak of backscattering light beats with the Brillouin laser, resulting in a frequency off sent of several hundreds of MHz. All the backscattering signals are detected by a 1.6GHz photodetector (Thorlabs PDB 480C) and recorded by a high speed data acquisition card (DAQ; Gage, 4GSa/s sampling rate) at 2GHz sampling rate.

*3.2 Vibration detection result*

The PZT tube used as the vibration source is at the position of 10164m of sensing fiber. The sensing fiber is wounded about 1.07m length on PZT, where the loop diameter and number of loops are 34mm and 10, respectively.

Fig. 5 show the φ-OTDR traces around PZT section (10155m to 10170m) when the PZT tube is applied by 100Hz and 1000Hz sinusoidal signal. Fig. 5(a) and 5(b) are superposition of ten consecutive Rayleigh backscattering traces recorded by DAQ, which show the amplitude changes caused by PZT. Due to the interference fading and the frequency shift of AOM, the traces are nearly zero at some place. Fig. 5(c) and 5(d) are normalized traces along different sampling periods over 50ms. The interference fading phenomenon is eliminated in these figures while vibration sections be emphasized. The phase change induced by PZT is clearly identified and the spatial resolution is ~3m which corresponding to 30 ns pulse.

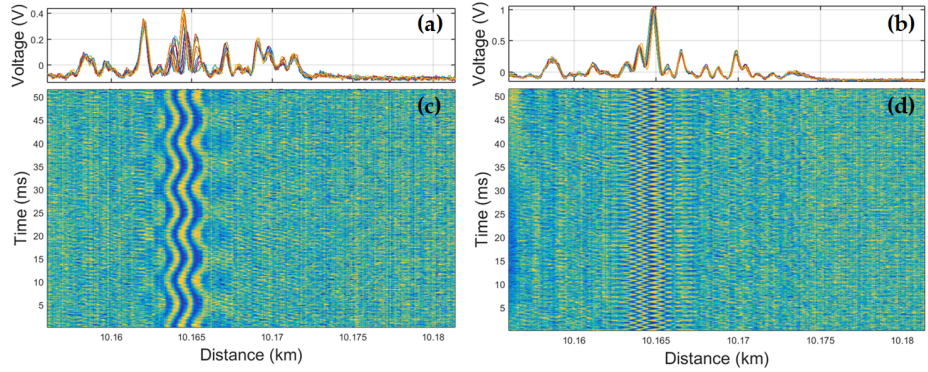

Fig. 5 φ-OTDR traces at end section of sensing fiber when PZT is driven by 100Hz (a) and 1KHz (b) sinusoidal signal, upper: superposition of ten traces, lower: consecutive traces within 50ms

In order to test the performance on different vibration frequency, 500Hz, 1KHz, 3KHz and 4.8KHz sinusoidal signals are applied to the PZT tube separately. The FFT transform spectrum at the vibration points, as shown in Fig. 6, demonstrates that the system has the capability to detect induced vibration along the fiber with up to 4.8KHz frequency response rang with a high SNR over 10dB.

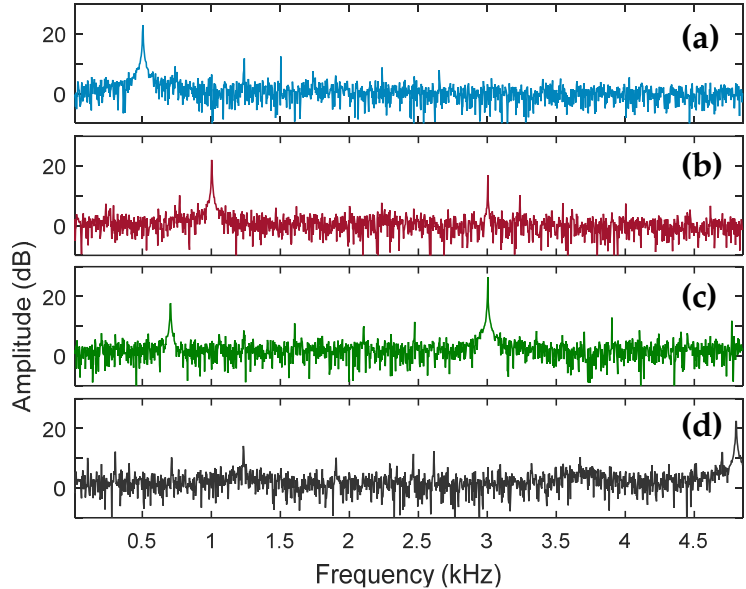

Fig. 6 (a)-(d) FFT transform spectrum of vibration point when applied 500Hz, 1KHz, 3KHz and 4.8KHz vibration.

### 3.3 Temperature and strain detection result

The acquired beat electrical signals of B-OTDR are analyzed by G-STFT method as mentioned in section 2. After 1000 cycles average, the spatial distribution of Brillouin scattering signal along the 10km fiber is shown in Fig. 7(a). BFS peak can be seen around 280MHz and attenuated along the fiber due to fiber losses. But it is high enough to be detected at the end of sensing fiber as shown in Fig. 7(b).

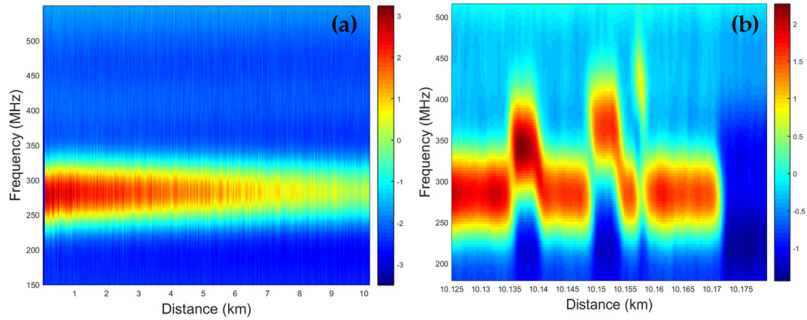

Fig. 7 (a) Spatial distribution of Brillouin scattering beat frequency signal along the 10km sensing fiber. (b) BFS of end section of the sensing fiber when applied temperate shift, strain and vibration simultaneously

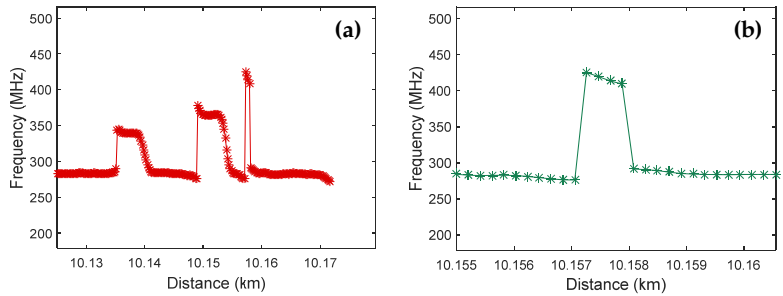

Fig. 8 (a) Brillouin frequency shift peaks of Fig. 9 extracted by Lorentzian nonlinear fitting. (b) enlarged frequency shift peak around the 80cm strain section in Fig. 5(a).

We also performed distributed temperate and strain measurement together with vibration measurement. Fig. 7(b) depicts the BFS spectrum distributions along the end section of sensing fiber when temperature, strain and vibration change. It demonstrates clearly that the frequency shift will occur at the temperate shift and strain regions, while no response at vibration point. This is because that the vibration induced disturbance is weak and it changing swiftly, which will be eliminated after averaging. The BFS peaks are extracted through windowed Lorentzian nonlinear fitting method shown in Fig. 8(a), through which we can easily calculate the frequency shift value and identify the temperature and strain shift location. Fig. 8(b) shows that the enlarged BFS peak graph around the 80cm strain section in Fig 8(a), indicating the spatial resolution is no less then 80cm. This resolution is an extremum for typical B-OTDR sensing due to the limit of phonon lifetime which is about 10ns in optical fiber [11].

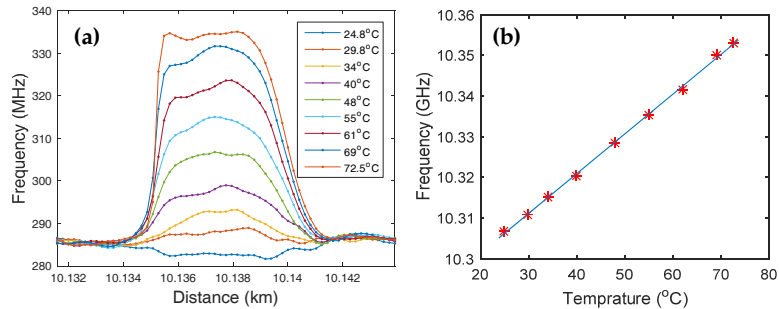

Fig. 9 (a) BFS peaks around water tank section under different temperature. (b) Relationship between the average Brillouin frequency and temperature

Brillouin temperature measurement is executed from 24.8°C to 72.5°C by changing the water temperature, which is calibrated by a 0.1°C resolution mercurial thermometer. Fig.9 (a) gives the BFS peaks around water tank section under different temperatures, which increases with water temperature. Fig. 9(b) shows the relationship between the average Brillouin frequency of the temperature shift section and the temperature measured by the mercurial thermometer. The average temperature-dependent coefficient of the fiber, calculated by linear fitting, is 0.9876MHz/°C.

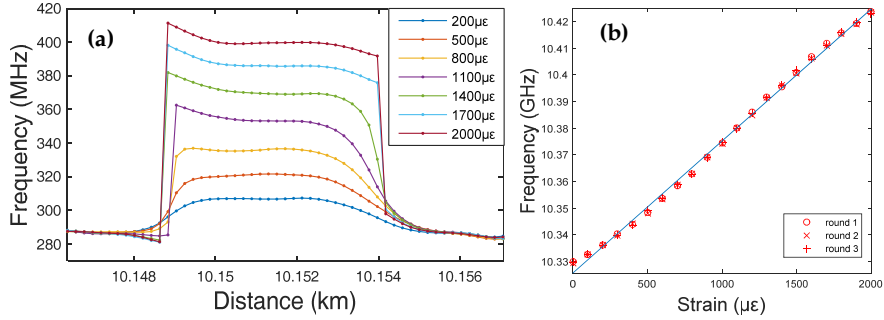

Fig. 10. (a) BFS peaks around 5m strain section under different strains. (d) Relationship between the average Brillouin frequency and strain

Shifting the micro-positioner of 5m fiber section by 0.5mm, we performed strain measurement with a range of 2000με. Fig.10(a) gives the BFS peaks around 5m section for different strains. Fig. 10(b) shows the relationship between the average Brillouin frequency and the strain applied to the fiber. It can be seen that the BFS shift within three test rounds is under 1MHz. The average strain-dependent coefficient of the fiber is 0.0495MHz/με.

*3.4 Discussions*

The experimental results indicate that the proposed method is capable for multi-parameter sensing, but there are many operating challenges when extending the performance of the system.

1). For distributed vibration sensing, the frequency response range here is 0~4.8 kHz for the limits of light pulse round trip time in the sensing fiber. While, in many application scenarios, like sound detection and metro railway monitoring, wide frequency response range is needed. Frequency division multiplexing method may be considered to enhance the frequency response of vibration detection.

Due to the arbitrary initial phase property of φ-OTDR, the amplitude of vibration, i.e. the accurate phase, is difficult to demodulate through the backscattering signals. A complex modulation of light pulse is promising to solve this problem. Inspired by the demodulation of MZI signals, 3×3 coupler and phase generated carrier techniques are also potential solutions.

2). For distributed temperature and strain sensing, our proposed solution has an 80cm spatial resolution which is an extremum for typical configuration due to the limit of phonon lifetime. A higher spatial resolution can be achieved when combining with the differential pulse-width pair method.

The sensing fiber used here is cross-sensitive for both temperate and strain. Introducing Raman signal demodulation into the system will eliminate this effect obviously, but it makes the setup complicated and costly because two additional sampling channels is needed for detecting Raman anti-Stokes (AS) and Stokes (S) signals. While in practical engineering projects, cross-sensitive problem can be partially overcame by using loose compact fiber or setting temperate reference fiber.

3) The 10 km sensing range value does not represent the limit and can be expanded by using longer optical pulses and, therefore, lower spatial resolution. The best trade-off between resolution and sensing range depends on the application. Besides, more than 100 km sensing range can be realized when amplifying the scattering light through Raman or EDFA amplifiers in sensing arm.

## 4. Conclusions

To conclude, a hybrid distributed multi-parameter fiber sensing system based on modulated pulses Φ/B-OTDR have been demonstrated to measure the vibration, temperature and strain with one photo detector and data acquisition channel. The probe sensing pulse is modulated into two kinds of profiles according to the fast changing property of vibration and the static property of temperature and strain, and then injected into sensing fiber sequentially with ratio of 100:1. In order to get a better spatial resolution, the pulses width are set to 30 ns and 8 ns separately for Phi-OTDR and B-OTDR sensing, matching with their different sensing principles. The modulation of the laser pulse width and intensity not only enables multi-parameter sensing, but also minimizes the nonlinear effect noise, which enhances the signal-to-noise ratio. Furthermore, Gaussian window short time Fourier transform method and weighted Lorentzian nonlinear fitting method are proceeded to demodulate $v_B$ in B-OTDR sensing, which a sum-meter level spatial resolution is achieved in temperature and strain detection eventually. The system can clearly identify up to 4.8kHz vibration at 10km distance along a standard single-mode fiber with 3m spatial resolution. And measurements of the distributed temperature and stress profile along the same fiber by 80 cm spatial resolution are also realized simultaneously.

## 5. Acknowledgments

This work is supported by the Project of Natural Science Foundation of China (Grant No: 61475029, 61377066, 61405020), The science fund for distinguished young scholars of Chongqing (Grant No: CSTC2014JCYJJQ40002) and The Project of Natural Science Foundation of Chongqing (Grant No: cstc2013jcyjA40029)